\documentclass[a4paper,11pt]{article}
\usepackage{graphicx}
\usepackage{multirow}
\usepackage{color}
\usepackage{latexsym}
\usepackage{amssymb}
\usepackage{amsfonts}
\usepackage{amsmath}
\usepackage{indentfirst}
\usepackage{slashbox,colortbl}
\newcommand{\cvd}{\hfill $\blacksquare$\bigskip}
\newtheorem{definition}{Definition}[section]

\newtheorem{proposition}{Proposition}[section]
\newtheorem{corollary}{Corollary}[section]

\author{Stefano Bilotta\thanks{Dipartimento di Matematica e Informatica ``U. Dini'', Universit\`a degli
Studi di Firenze, Viale
 G.B. Morgagni 65, 50134 Firenze, Italy. {
 \tt \ stefano.bilotta@unifi.it}}}

\title{Variable-length Non-overlapping Codes}

\date{}

\begin{document}

\maketitle

\begin{abstract}
We define a variable-length code having the property that no (non-empty) prefix of each its codeword is a suffix of any other one, and vice versa. This kind of code can be seen as an extension of two well-known codes in literature, called respectively \emph{fix-free code} and \emph{non-overlapping code}. In this paper, some constructive algorithms for such codes are presented as well as numerical results about their cardinality.
\end{abstract}


\section{Introduction}
A set of codewords is \emph{fix-free} (sometimes called \emph{affix code}, \emph{biprefix code} or \emph{never-self-synchronizing code}) if it is both prefix-free and suffix-free: no codeword in the set is a prefix or a suffix of any other.
In a fix-free code, any finite sequence of codewords can be decoded in both directions, thus reducing the decoding time and error propagation. This kind of variable-length codes, introduced six decades ago by Sch\"utzenberger \cite{schutz} and by Gilbert and Moore \cite{gilbert}, have several applications, for example they have been recently used in certain international video compression standards for robustness to channel errors \cite{bauer,gao}.


A string $v$ over a finite alphabet $\Sigma$ is called \emph{bifix-free} (or equivalently \emph{unbordered} or \emph{self non-overlapping}) if it does not contain proper prefixes which are also proper suffixes. In other words, a string $v \in \Sigma^*$ is bifix-free if it cannot be factorized as $v=uwu$ with $u \in \Sigma^+$ and $w \in \Sigma^*$, denoting by $\Sigma^*$
the set of all finite strings over $\Sigma$ and by $\Sigma^+$ the subset of $\Sigma^*$ excluding the \emph{empty string}.

Nielsen in \cite{nielsen} provided the set $X \subset \Sigma^n$ of all $n$-length bifix-free strings by means of a recursive construction. More recently, several researches \cite{bajic,motzkin,bilo,black,singa} have been conducted in order to define particular subsets of $X$ constituted by \emph{cross-bifix-free} (or \emph{non-overlapping}) strings: two $n$-length strings $v,v' \in X$ are called cross-bifix-free if any non-empty proper prefix of $v$ is different from any non-empty proper suffix of $v'$, and vice versa.

A set of $n$-length strings over a given alphabet is said to be a cross-bifix-free set (also known as \emph{cross-bifix-free code} or \emph{non-overlapping code}) if and only if any two elements in the set are cross-bifix-free.

Cross-bifix-free codes are involved in the study of frame synchronization \cite{bajsto,delind} which is an essential
requirement in a digital communication systems to establish and maintain a connection between a
transmitter and a receiver. In particular, this kind of codes have the strong property that an error in a codeword or in a state of a certain decoding automaton does not propagate into incorrect decoding of subsequent codewords.

The problem of determining such sets is also related
to several other scientific applications, for instance, in the theory of formal languages for the generation and the enumeration of particular binary words avoiding a set of given factors \cite{acta}, and in the DNA-based storage systems for the characterization of DNA mutually uncorrelated codes \cite{Yadzi}.

In this paper, we are interested in the study of a set of codewords which can be seen an extension of cross-bifix-free and fix-free codes.
Since the theory of such codes are widely used in several
fields of applications, we are expected that the present extension could have the same usefulness and it could constitute a starting point
for a fruitful and intriguing theory. To achieve this purpose, we have to extend the concept of cross-bifix-free set to strings having not necessarily the same length, taking into account that no string in the set is a prefix or a suffix of any other.

In the next section we define a set of codewords over a binary alphabet satisfying such constraints. In Section 3 we analyze its cardinality considering some similarities with the well-known \emph{$k$-generalized Fibonacci numbers} and we give a rational generating function for the enumeration of our set according to the length of its codewords. In section 4 we generalize the study to codes having greater cardinality than the above one, and we determine a code, by using the well-known \emph{Dyck words}, having the property to be non-expandable.

\section{A binary cross-fix-free code}
We introduce the following definition which is based on the natural extension of the notion of cross-bifix-free or non-overlapping strings to strings having variable-length.

\begin{definition}\label{defcode}
A variable-length code is said to be a \emph{cross-fix-free code} or \emph{variable-length non-overlapping code} if and only if
for any two codewords in the set no (non-empty) prefix of the first one is a suffix of the second one, and vice versa.
\end{definition}

\noindent Please notice that in the above definition if the prefix of a codeword coincides exclusively with the codeword itself, then the resulting code is fix-free.\\

In order to construct one among all of the possible cross-fix-free sets over a binary alphabet, we first consider the following set of $n$-length strings.

\begin{definition}\label{defi1}
For any fixed $k$, with $3 \leq k \leq \lfloor n/2\rfloor -2$, we denote by $V_n^{(k)}$ the set of all $n$-length binary strings $v=v_1v_2\ldots v_n$ that satisfy the following properties:
\begin{itemize}
\item $v_1=v_2=\ldots =v_k=1$;
\item $v_{n-k+1}=v_{n-k+2}=\ldots=v_n=0$
\item the subsequence $v_{k+1}v_{k+2}\ldots v_{n-k}=0w1$ can contain neither $k$ consecutive 0's nor $k$ consecutive 1's.
\end{itemize}
\end{definition}

In other words, for any fixed $k$, the set $V_n^{(k)}$ contains the $n$-length binary strings starting with a prefix of $k$ consecutive 1's (denoted by $1^k$), ending by a suffix of $k$ consecutive 0's
(denoted by $0^k$), and the factor which is included between them begins with 0, ends with 1 and avoids both $0^k$ and $1^k$.

For example,
$$V_{11}^{(3)}=\left\{11101101000,11101001000,11101011000,11100101000\right\}.$$

We note that, for $k=2$ and $n$ odd, the set $V_n^{(k)}$ cannot be defined since the factor $0w1$ cannot avoid both $00$ and $11$.

\begin{proposition}\label{cross-bifix}
For any fixed $k$, with $3 \leq k \leq \lfloor n/2\rfloor -2$, $V_n^{(k)}$ is a cross-bifix-free code.
\end{proposition}

\emph{Proof.} \quad It consists of two distinguished steps. First we prove that each $v \in V_n^{(k)}$ is bifix-free, then we show that $V_n^{(k)}$ is a cross-bifix-free code. Each $v \in V_n^{(k)}$ can be written as $v=\alpha w \beta$ where $\alpha, \beta$ are necessarily non-empty strings while $w$ can also be empty. Denoting by $|\alpha|$ the length of the string $\alpha$, we consider two cases. If $|\alpha|=|\beta|=i \leq k$, then $\alpha=1^i \neq 0^i=\beta$, so $v$ is bifix-free. If $|\alpha|=|\beta| > k$, then $\alpha=1^k0\nu$ where $\nu$ can be empty and $0\nu$ avoids both $0^k$ and $1^k$, and $\beta = \nu'10^k$ where $|\nu'|=|\nu|$ and $\nu'0$ avoids both $0^k$ and $1^k$. Supposing ad absurdum that $v$ is not bifix-free, if $|\nu|=|\nu'| \geq k-1$, then $\nu'1$ would contain $1^k$ against the hypothesis $v \in V_n^{(k)}$, otherwise, if $|\nu|=|\nu'| \leq k-2$, then there would be a mismatch between $v_{|\nu|+2}$ and $v_{n-k+1}$ which are equal to 1 and 0, respectively. Therefore, $v$ is bifix-free.

The proof that $v$ and $v'$ are cross-bifix-free for each $v, v' \in V_n^{(k)}$ is quite analogous to the one just illustrated, considering $v=\alpha w$ and $v'= w' \beta'$ and comparing the proper prefix $\alpha$ of $v$ and the proper suffix $\beta'$ of $v'$ when $|\alpha|=|\beta'|$.
\cvd

In order to obtain a set of binary string having different length we are going to consider the following union
\begin{equation}\label{unione}
\mathcal V^{(k)}=\bigcup_{i\geq 2k+2} V_i^{(k)}
\end{equation}

\noindent
and $\mathcal V^{(k)}(n)$ denotes the subset of $\mathcal V^{(k)}$ containing binary strings having length at most $n$.

In Table \ref{vk} we list the first elements of the set $\mathcal V^{(3)}$ according to the length of the strings (in particular, the set $\mathcal V^{(3)}(13)$ is shown).

\begin{table}[h!]
\begin{center}
{\footnotesize
\begin{tabular}{lc}
n=8 & 11101000\\
\hline
n=9 & 111011000\\
&111001000\\
\hline
n=10 & 1110101000\\
&1110011000\\
\hline
n=11 & 11101101000\\
&11101001000\\
&11101011000\\
&11100101000\\
\hline
n=12 & 111010011000\\
&111011011000\\
&111011001000\\
&111010101000\\
&111010011000\\
&111001101000\\
&111001001000\\
\hline
n=13 &1110011011000\\
&1110010011000\\
&1110011001000\\
&1110010101000\\
&1110110011000\\
&1110110101000\\
&1110100101000\\
&1110101101000\\
&1110101011000\\
&1110101001000\\
\hline
&\vdots
\end{tabular}
}
\caption{First elements of $\mathcal V^{(3)}$.} \label{vk}
\end{center}
\end{table}

\begin{proposition}\label{singolo}
For any fixed $k\geq 3$, $\mathcal V^{(k)}$ is a cross-fix-free code.
\end{proposition}

\emph{Proof.} \quad From Proposition \ref{cross-bifix} it follows that $V_i^{(k)}$ is a cross-bifix-free code, for each $i \geq 2k+2$. To complete the proof we have to show that $V_i^{(k)} \cup V_j^{(k)}$ is a cross-fix-free code, for each $i,j \geq 2k+2$ with $i<j$.
First of all, each $v \in V_i^{(k)}$ can be written as $v=\alpha w$ and each $v' \in V_j^{(k)}$ can be written as $v'=w' \beta'$, for some $w,w',\alpha, \beta' \in \{0,1\}^+$. By using the same argument of the proof of Proposition \ref{cross-bifix}, it is not difficult to realize that $\alpha \neq \beta'$ for each $\alpha,\beta'$ such that $|\alpha|=|\beta'|$, so no (non-empty) proper prefix of $v$ is a suffix of $v'$, and vice versa, for each $v \in V_i^{(k)}$ and $v'\in V_j^{(k)}$.

\begin{figure}[htb]
\begin{center}
\includegraphics[scale=0.6]{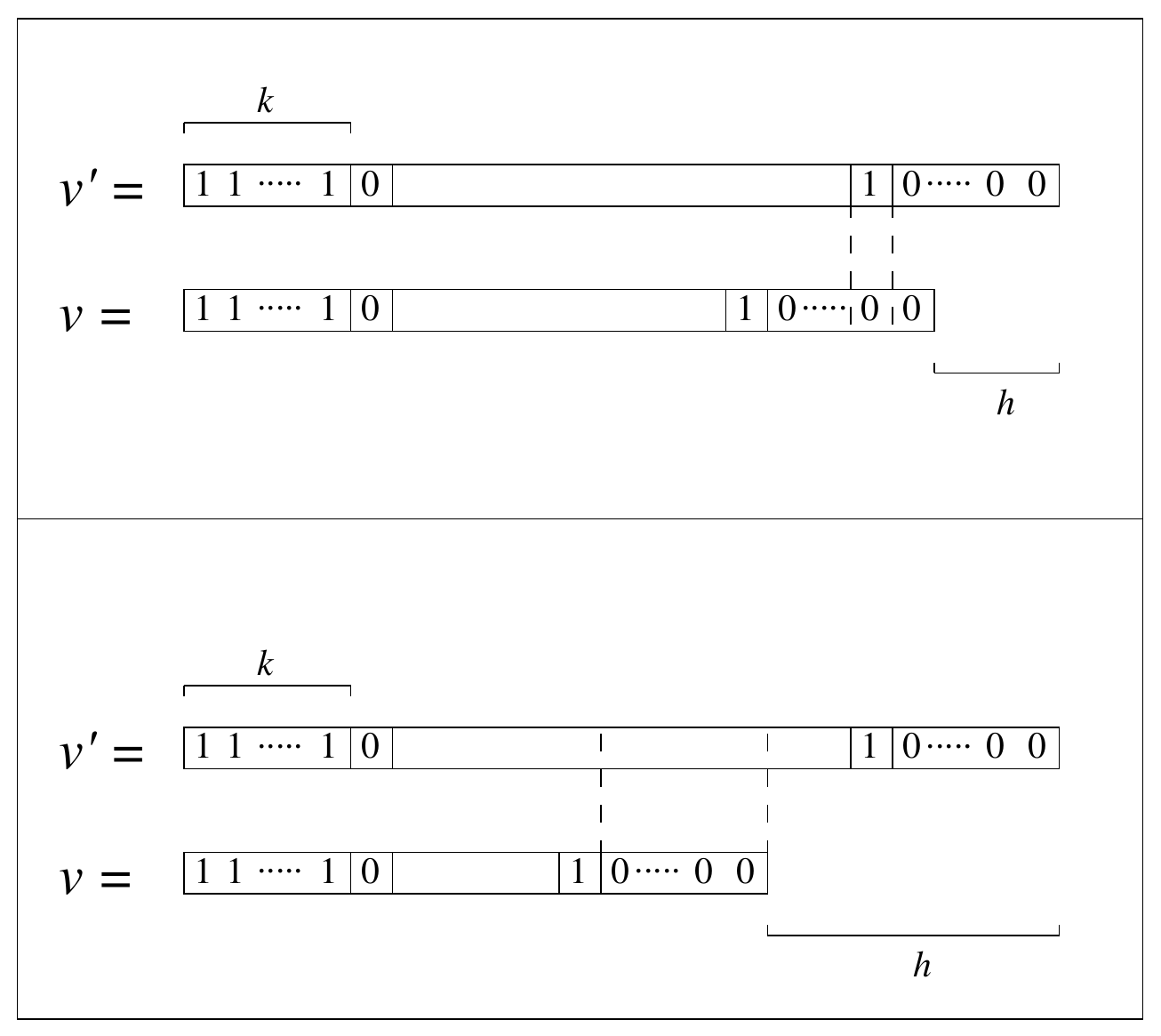} \caption{The strings $v' \in V^{(k)}_{i+h}$ and $v \in V^{(k)}_i$. In the top $|v'|-|v|= h \leq k$, in the bottom $|v'|-|v|= h >k$.}\label{prova}
\end{center}
\end{figure}

Now, we have to prove that $V_i^{(k)} \cup V_j^{(k)}$ is also fix-free, that is, no codeword in the set is a prefix or a suffix of any other. Suppose ad absurdum that $v'= v \gamma$ with $\gamma \in \{0,1\}^+$ and $|\gamma|=j-i=h>0$. Then, we have two cases: $h \leq k$ and $h>k$. In the former we have that $v'_{j-k}=v_{i+h-k}$ but $v'_{j-k}=1$ and $v_{i+h-k}=0$, leading a contradiction (see the top of Figure \ref{prova}). In the latter we have that $v'_{j-h-k+1}v'_{j-h-k+2}\ldots v'_{j-h}=v_{i-k+1}v_{i-k+2}\ldots v_i$ but $v_{i-k+1}v_{i-k+2}\ldots v_i =0^k$ against the hypothesis $v' \in V_j^{(k)}$ (see the bottom side of Figure \ref{prova}). Analogously, $v'\neq \gamma v$ for any $\gamma \in \{0,1\}^+$, hence $\mathcal V^{(k)}$ is a cross-fix-free code.

\cvd

Following the just described argument in the above proof, it is easy to see that $\mathcal V^{(k)}$ has also another strong property: each codeword cannot be a factor of another codeword in the set.

\begin{corollary}\label{nofactor}
For each $v,v' \in \mathcal V^{(k)}$, $v$ cannot be a factor of $v'$, for any $k \geq 3$.
\end{corollary}

\emph{Proof.} \quad Let be $v \in V_i^{(k)}$ and $v'\in V_j^{(k)}$, with $i<j$. We have to suppose ad absurdum that $v'=\gamma v \gamma'$ with $\gamma,\gamma' \in \{0,1\}^+$. If $|\gamma|=h \leq k$, then $v'_{k+1}=v_{k+1-h}$ but $v'_{k+1}=0$ and $v_{k+1-h}=1$, leading a contradiction. If $|\gamma|=h>k$, then $v'_{h+1}v'_{h+2}\ldots v'_{h+k}=v_1v_2\ldots v_k$ but $v_1v_2\ldots v_k=1^k$ against the hypothesis $v' \in V_j^{(k)}.$
\cvd

Please notice that, all the results presented up to now are still valid if we consider the set $\mathcal V^{(k)} \cup \{1^k0^k\}$, for any $k \geq 3$.

\section{The enumeration of $\mathcal V^{(k)}(n)$}\label{number}

We denote by $R_{\ell}(0^k,1^k)$, with $\ell\geq 1$, the set of $\ell$-length binary strings starting with $0$, ending with $1$ and avoiding $k$ consecutive $0$'s and $k$ consecutive $1$'s. We consider that $R_0(0^k,1^k) =\{\lambda\}$, where $\lambda$ is the empty string, and we indicate with $r_{\ell}^{(k)}$ the
cardinality of $R_{\ell}(0^k,1^k)$. From Definition \ref{defi1}, for any fixed $k \geq 3$, it is straightforward that the cardinality of $V^{(k)}_n$, denoted by $|V^{(k)}_n|$, results to be

\begin{equation*}
|V^{(k)}_n|=r_{n-2k}^{(k)}, \quad n \geq 2k+2
\end{equation*}

\noindent thus,

\begin{equation}\label{partialset}
|\mathcal V^{(k)}(n)|=\left|\bigcup_{i=2k+2}^n V^{(k)}_i\right|= \sum_{i=2k+2}^n r_{i-2k}^{(k)}= \sum_{\ell=2}^{n-2k} r_{\ell}^{(k)}.
\end{equation}

The sequence $r_{\ell}^{(k)}$ can be expressed in terms of the well-known $k$-\emph{generalized Fibonacci numbers}. We recall that, the $k$-generalized Fibonacci numbers $\left\{\text{f}_{\ell}^{(k)}\right\}_{{\ell}\geq 0}$ can be defined as (see \cite{knuth})

\begin{equation*}
\text{f}_{\ell}^{(k)}=\left \{
\begin{array}{ll}
0 & \mbox{if} \ 0 \leq {\ell} < k-1\\
\\
1 & \mbox{if} \ \ell=k-1\\
\\
\displaystyle \sum_{i=1}^k \text{f}_{{\ell}-i}^{(k)} & \mbox{if} \ {\ell} \geq k\ .
\end{array}\right.
\end{equation*}

For our purpose, we are interested in the sequence $\left\{\text{f}^{(k)}_{\ell+k}\displaystyle\right\}_{\ell \geq 0}$ counting the number of ${\ell}$-length binary strings avoiding $0^k$. Posing $f_{\ell}^{(k)}=\text{f}_{\ell+k}^{(k)}$ we have

\begin{equation}\label{kbonaccistring}
f_{\ell}^{(k)}=\left \{
\begin{array}{ll}
2^\ell & \mbox{if} \ 0 \leq {\ell} < k-1\\
\\
\displaystyle \sum_{i=1}^k f_{{\ell}-i}^{(k)} & \mbox{if} \ {\ell} \geq k\ .
\end{array}\right.
\end{equation}

\bigskip

It can be proved, by induction \cite{matrix}, that

\begin{equation}\label{rlfib}
r_{\ell}^{(k)}=\left \{
\begin{array}{ll}
1 & \mbox{if} \ {\ell} = 0\\
\\
\displaystyle\frac{f^{(k-1)}_{{\ell}-1} + d^{(k)}_{\ell}}{2} & \mbox{if} \ {\ell} \geq 1,
\end{array}
\right.
\end{equation}

\noindent where

\begin{equation*}
d^{(k)}_{\ell}=
\left\{
\begin{array}{rl}
1& \mbox{if}\ ({\ell}\ \mbox{mod}\ k)=0\\
\\
-1& \mbox{if}\ ({\ell}\ \mbox{mod}\ k)=1\\
\\
0& \mbox{if}\ ({\ell}\ \mbox{mod}\ k)\geq 2\ .\\
\end{array}
\right.
\end{equation*}

In Table \ref{rn} we list the first numbers of the recurrence $r_{\ell}^{(k)}$ for some fixed values of $k$.

\begin{table}[h!]
\begin{center}
{\footnotesize
\begin{tabular}{|c|rrrrrr|}
\hline
\backslashbox{${\ell}$}{$k$} & 3 & 4 & 5 & 6 & 7 & 8\\
\hline
0 & 1 & 1 & 1 & 1 & 1 & 1\\
1 & 0 & 0 & 0 & 0&0&0\\
2 & 1 & 1 & 1 & 1&1&1\\
3 & 2 & 2 & 2 & 2&2&2\\
4 & 2 & 4 & 4 & 4&4&4 \\
5 & 4 & 6 &  8&8&8&8  \\
6 & 7 & 12 & 14 & 16 & 16 & 16\\
7&10&22&28&30&32&32\\
8&17&41&54&60&62&64\\
9&28&74&104&118&124&126\\
10&44&137&201&232&246&252\\
11&72&252&386&456&488&502\\
12&117&464&745&897&968&1000\\
13&188&852&1436&1762&1920&1992\\
14&305&1568&2768&3465&3809&3968\\
15&494&2884&5336&6812&7554&7904\\

\hline\end{tabular}
}
\caption{Sequences $r_{\ell}^{(k)}$ for some fixed values of $k$.} \label{rn}
\end{center}
\end{table}

From the equations (\ref{partialset}) and (\ref{rlfib}) it follows that

\begin{equation}\label{presumfib}
|\mathcal V^{(k)}(n)|= \displaystyle \sum_{\ell=2}^{n-2k} \left(\frac{f^{(k-1)}_{{\ell}-1} + d^{(k)}_{\ell}}{2} \right).
\end{equation}

\noindent The sequence
$\left\{d_{\ell}^{(k)}\right\}_{\ell \geq 0}=\Big\{1,-1,
\underbrace{0,0,\ldots,0}_{k-2},1,-1,
\underbrace{0,0,\ldots,0}_{k-2},1,-1,0,\ldots\Big\}
$
is such that
$\sum_{i=0}^{k-1}d_{{\ell}+i}^{(k)}=0$, for each ${\ell}\geq 0$. Note that $d_\ell^{(k)}=1$ when $\ell=0,k,2k,3k,\ldots$, and the sum $\displaystyle \sum_{\ell=2}^{n-2k}  d^{(k)}_{\ell} =1$ if its $(n-2k)$-th term is equal to 1. Therefore,
$\displaystyle \sum_{\ell=2}^{n-2k}  d^{(k)}_{\ell} = \delta_n^{(k)} = 1$ if $(n \ \mbox{mod} \ k)= 0$, and otherwise it is 0. So, equation (\ref{presumfib}) becomes

\begin{equation}\label{sumfib}
|\mathcal V^{(k)}(n)|= \frac{\left(\displaystyle\sum_{\ell=2}^{n-2k} f^{(k-1)}_{{\ell}-1}\right) + \delta_n^{(k)}}{2}.
\end{equation}

In order to further simplify the calculus of (\ref{sumfib}) we are going to consider a general study of the partial sums of $\left\{f_\ell^{(k)}\right\}_{\ell\geq0}$, for any fixed $k \geq 3$.

Denoting by
$S^{(k)}(n) = f_0^{(k)}+ f_1^{(k)}+\ldots+f_n^{(k)}=\displaystyle \sum_{\ell=0}^n f_{\ell}^{(k)}$, it is well-known that $\left\{S^{(2)}(n)\right\}_{n\geq0}=\left\{1,3,6,11,19,32,53, \ldots\right\}$ can be expressed as $S^{(2)}(n)=f_{n+2}^{(2)}-2$ for $n \geq 0$ (see sequence A001911 in The On-line Encyclopedia of Integer Sequences), being $\left\{f_\ell^{(2)}\right\}_{\ell \geq 0}=\left\{1,2,3,5,8,13,21,34,55,\ldots\right\}$.

Now, we first consider a similar formulation for $S^{(3)}(n)$ by means of the following proposition and then we generalize it to any fixed $k>3$.

\begin{proposition}\label{sommatre}
We have that

\begin{displaymath}
S^{(3)}(n)=\left\{
\begin{array}{ll}
2^{n+1}-1 & \mbox{if} \ 0 \leq n \leq 2\\
\\
\displaystyle\frac{f_{n+2}^{(3)}-2^3+1+f_1^{(3)}+f_n^{(3)}}{2} +f_0^{(3)} & \mbox{if} \ n \geq 3.
\end{array}
\right.
\end{displaymath}
\end{proposition}

\emph{Proof.} \quad
We only consider the case $n \geq3$. From (\ref{kbonaccistring}) we have that $f_3^{(3)}= 2^3-1$. If the term $\left ( f_1^{(3)} + f_2^{(3)}\right)$ is added to both the sides of the previous equation, being $f_4^{(3)}=f_3^{(3)}+f_2^{(3)}+f_1^{(3)}$, then

$$ f_4^{(3)} = 2^3 - 1 +f_1^{(3)}+f_2^{(3)}.$$

Analogously, if the term $\left ( f_2^{(3)} + f_3^{(3)}\right)$ is added to both the sides of the last equation, then we have

$$f_5^{(3)}= 2^3-1 +f_1^{(3)}+2f_2^{(3)}+f_3^{(3)}.$$

%

At the $n$-th step, the described procedure leads to

$$f_{n+2}^{(3)}=2^3-1+f_1^{(3)}+2(f_2^{(3)}+f_3^{(3)}+\ldots+f_{n-1}^{(3)})+f_n^{(3)}.$$

In the right side of the above equation all the terms $f_i^{(3)}$ appear, with $i=1,2,\ldots,n$. Thus, we can rewrite it in terms of $S^{(3)}(n)-f_0^{(3)}$ as

$$f_{n+2}^{(3)}= 2^3 - 1 + 2(S^{(3)}(n)-f_0^{(3)}) -f_1^{(3)}-f_n^{(3)},$$

and the thesis follows.\cvd


The result in Proposition \ref{sommatre} can be generalized to $k \geq 3$ by means of the following proposition.

\begin{proposition} For any fixed $k \geq 3$, we have that
\begin{displaymath}
S^{(k)}(n)=\left\{
\begin{array}{ll}
2^{n+1}-1 & \mbox{if} \ 0 \leq n \leq k-1\\
\\
\displaystyle\frac{f_{n+2}^{(k)}-2^k+1+\displaystyle \sum_{i=0}^{k-3}\left( (k-2-i)(f_{i+1}^{(k)}+f_{n-i}^{(k)})\right)}{k-1} +f_0^{(k)} & \mbox{if} \ n \geq k.
\end{array}
\right.
\end{displaymath}
\end{proposition}

\emph{Proof.} \quad We can proceed by induction for $n \geq k$. If $n=k$, then we have

\begin{displaymath}
\begin{array}{rl}
S^{(k)}(k)=& \displaystyle\frac{f_{k+2}^{(k)} -2^k+1 + (k-2)(f_1^{(k)} + f_k^{(k)}) + (k-3)(f_2^{(k)}+ f_{k-1}^{(k)})}{k-1}+\\
\\
&+\displaystyle\frac{(k-4)(f_3^{(k)}+f_{k-2}^{(k)})+\ldots+ (f_{k-2}^{(k)}+f_3^{(k)})}{k-1} + f_0^{(k)}\\
\\
=&\displaystyle\frac{f_{k+2}^{(k)} -f_k^{(k)} + (k-2)(f_1^{(k)} + f_k^{(k)})}{k-1}+ \\
\\
&+\displaystyle\frac{(k-3)(f_2^{(k)}+f_3^{(k)}+ \ldots +f_{k-2}^{(k)}+ f_{k-1}^{(k)})}{k-1} + f_0^{(k)}.
\end{array}
\end{displaymath}

Since $f_{k+2}^{(k)}-f_k^{(k)}=2(f_2^{(k)}+f_3^{(k)}+ \ldots +f_{k-2}^{(k)}+ f_{k-1}^{(k)}) + (f_1^{(k)} + f_k^{(k)})$, then

$$S^{(k)}(k)=\sum_{i=0}^{k}f_i^{(k)}$$ and the base case $n=k$ is verified.

By inductive hypothesis we suppose that, for $s < n$, $$S^{(k)}(s)=\displaystyle\frac{f_{s+2}^{(k)} - 2^k +1+ \displaystyle\sum_{i=0}^{k-3}\left ( (k-2-i)(f_{i+1}^{(k)}+ f_{s-i}^{(k)})\right)}{k-1} + f_0^{(k)}.$$

Since $S^{(k)}(n)=S^{(k)}(n-1)+f_n^{(k)}$, then we have that

\begin{displaymath}
\begin{array}{rl}
S^{(k)}(n)=& \displaystyle\frac{f_{n+1}^{(k)} - 2^k +1+ \displaystyle\sum_{i=0}^{k-3}\left ( (k-2-i)(f_{i+1}^{(k)}+ f_{n-1-i}^{(k)})\right)}{k-1} + f_0^{(k)} +f_n^{(k)}\\
\\
=& \displaystyle\frac{f_{n+1}^{(k)} - 2^k +1+ \displaystyle\sum_{i=0}^{k-3}\left ( (k-2-i)f_{i+1}^{(k)}\right)}{k-1}+\\
\\
& +\displaystyle\frac{(k-1)f_n^{(k)}+ (k-2)f_{n-1}^{(k)}+\ldots + f_{n+2-k}^{(k)}}{k-1}+ f_0^{(k)}.\\
\end{array}
\end{displaymath}

Since $f_{n+2}^{(k)}=f_{n+1}^{(k)}+f_n^{(k)}+ \ldots + f_{n+2-k}^{(k)}$, then the thesis follows. \cvd

From equation (\ref{sumfib}) it follows that, for any $n \geq 2k+2$

$$ |\mathcal V^{(k)}(n)|= \displaystyle \frac{S^{(k-1)}(n -2k -1) - f_0^{(k-1)}+ \delta_n^{(k)}}{2}, $$

\noindent recalling that $f_0^{(k-1)}=1$, for any $k \geq 3$, and $\delta_n^{(k)}=1$ if $(n \ \mbox{mod} \ k)=0$, and otherwise it is 0.
For example, when $k=3$, we have

$$ |\mathcal V^{(3)}(n)|= \displaystyle \frac{f_{n-5}^{(2)}-3 + \delta_n^{(3)}}{2}, \ \ n \geq 8. $$

\subsection{Generating functions}

The rational generating function $f^{(k)}(x)$ of the sequence $\left\{f_{\ell}^{(k)}\right\}_{{\ell} \geq 0}$ (see A000045 for $k=2$, A000073 for $k=3$, A000078 for $k=4$ in The On-line Encyclopedia of Integer Sequences) is given by
\begin{equation}\label{knacci}
f^{(k)}(x)=\frac{1+\displaystyle\sum_{i=1}^{k-1}x^i}{1-\displaystyle\sum_{i=1}^{k}x^i} \ .
\end{equation}

The rational generating function $d^{(k)}(x)$ of the sequence $\left\{d_{\ell}^{(k)}\right\}_{{\ell}\geq 0}$
(see sequences A049347 for $k=3$, A219977 for $k=4$, in The On-line Encyclopedia of Integer Sequences) is given by
$$
d^{(k)}(x)= \frac{1}{1+\displaystyle\sum_{i=1}^{k-1}x^i} \ ,
$$

\noindent
and, from equation (\ref{rlfib}), the generating function $r^{(k)}(x)$ of the sequence $\left\{r_{\ell}^{(k)}\right\}_{{\ell}\geq 2}$ can be easily obtained as
$$
r^{(k)}(x)= \frac{x f^{(k-1)}(x) + d^{(k)}(x)-1}{2}\ .
$$

Given a generating function $a(x)=\displaystyle \sum_{n \geq 0} a_n x^n$, it is well-known that the generating function $s(x)$ of the sequence $\left\{\displaystyle\sum_{i=0}^n a_i\right\}_{n\geq0}$ is given by $s(x)=\displaystyle\frac{a(x)}{1-x}$, hence, from equation (\ref{partialset}), it is easy to deduce that the generating function $\mathcal V^{(k)}(x)$ of the sequence $\displaystyle\left\{ |\mathcal V^{(k)}(n)|\right\}_{n\geq2k+2}$, according to the length of its codewords, is

$$\mathcal V^{(k)}(x)=\frac{x^{2k}r^{(k)}(x)}{1-x}=\frac{x^{2k}(x-x^k)^2}{(1-x)(1-x^k)(1-2x+x^k)}.$$

\noindent In Table \ref{vnsize} we list the first numbers of $|\mathcal V^{(k)}(n)|$ for some fixed values of $k$.

\begin{table}[h!]
\begin{center}
{\footnotesize
\begin{tabular}{|c|rrrrrrrrr|}
\hline
\backslashbox{${n}$}{$k$} & 3 & 4 & 5 & 6 & 7 & 8 & 9 & 10 & 11 \\
\hline
8 & 1 &  &  &  & & &&&\\
9 & 3 &  &  &  & &&&&\\
10 & 5 & 1 &  &  & &&&&\\
11 & 9 & 3 &  &  & &&&&\\
12 & 16 & 7 & 1 &  & &&&&\\
13 & 26 & 13 & 3 &  & &&&&\\
14 & 43 & 25 & 7 & 1 & &&&&\\
15 & 71 & 47 & 15 & 3 & &&&&\\
16 & 115 & 88 & 29 & 7 & 1 &&&&\\
17 & 187 & 162 & 57 & 15 & 3 &&&&\\
18 & 304 & 299 & 111 & 31 & 7 & 1&&&\\
19 & 492 & 551 & 215 & 61 & 15 & 3 &&&\\
20 & 797 & 1015 & 416 & 121 & 31 & 7 & 1 &&\\
21 & 1291 & 1867 & 802 & 239 & 63 & 15 & 3&&\\
22 & 2089 & 3435 & 1547 & 471 & 125 & 31 & 7 & 1&\\
23 & 3381 & 6319 & 2983 & 927 & 249 & 63 & 15 & 3&\\
24 & 5472 & 11624 & 5751 & 1824 & 495 & 127 & 31 & 7 & 1\\

\hline\end{tabular}
}
\caption{$|\mathcal V^{(k)}(n)|$ for some fixed values of $k$.} \label{vnsize}
\end{center}
\end{table}

\section{Further developments}

In this section we generalize the study of Section 2 in order to obtain codes, having greater cardinality than the previous one. We recall that, the cross-fix-free code $\mathcal V^{(k)}$ admits also the strong property that no codeword is a factor of any other, for any fixed $k \geq3$ (see Corollary \ref{nofactor}). In the case, we refer to this kind of codes as \emph{strong cross-fix-free codes} or \emph{strong variable-length non-overlapping codes}. However, Definition \ref{defcode} does not require that a such property have to be verified, so we can distinguish the strong cross-fix-free codes from the other ones. In particular, in this section we are going to construct the other ones.

A first attempt in this direction can be constituted by the union of $\mathcal V^{(k)}$, for each $k \geq 3$. Then, we have the following proposition.

\begin{proposition}
$\mathcal W= \displaystyle\bigcup_{k\geq 3} \mathcal V^{(k)}$ is a cross-fix-free code.
\end{proposition}

\emph{Proof.} \quad Proposition \ref{singolo} states that $\mathcal V^{(k)}$ is a cross-fix-free code, for each $k \geq 3$. Then, we have to prove that $\mathcal V^{(k)} \cup \mathcal V^{(k')}$ is a cross-fix-free code, for each $k,k'\geq 3$. Fixed $k,k'$ and suppose that $k<k'$, the first step of the proof consists in showing that $\mathcal V^{(k)} \cup \mathcal V^{(k')}$ is a fix-free code. Let be $v \in \mathcal V^{(k)}$ and $v' \in \mathcal V^{(k')}$, it is easily to realize that $v_{k+1}=0$ while $v'_{k+1}=1$, so that $v$ cannot be a prefix of $v'$, and vice versa. Similarly, $v$ cannot be a suffix of $v'$ and vice versa. To complete the proof we have to show that each (non-empty) proper prefix of $v$ cannot be a suffix of $v'$, and vice versa. The different values of $k$ and $k'$ do not change the argument used in the proof of Proposition \ref{singolo}, so we can conclude that $\mathcal W$ is a cross-fix-free code.
\cvd

Note that, $\mathcal W$ contains codewords which are factors of some others. For example, $v'=111110\textbf{11101000}0100000 \in \mathcal V^{(5)}$ contains $v=11101000 \in \mathcal V^{(3)}$.

Denoting by $\mathcal W (n) = \displaystyle\bigcup_{k=3}^{\lfloor\frac{n-2}{2}\rfloor} \mathcal V^{k}(n)$ and using the results in Section 3, the first numbers of $|\mathcal W(n)|$ are:
$$1,3,6,12,24,42,76,136,240,424,753,1337,2388,4280,7706,13940,25332,$$
\noindent for $n=8,9,\ldots,24$, respectively.

In Table \ref{w} we list the first elements of the set $\mathcal W$ according to the length of the strings (in particular, the set $\mathcal W(12)$ is shown).

\begin{table}[h!]
\begin{center}
{\footnotesize
\begin{tabular}{lc}
n=8 & 11101000\\
\hline
n=9 & 111011000\\
&111001000\\
\hline
n=10 & 1110101000\\
&1110011000\\
&1111010000\\
\hline
n=11 & 11101101000\\
&11101001000\\
&11101011000\\
&11100101000\\
&11110010000\\
&11110110000\\
\hline
n=12 & 111010011000\\
&111011011000\\
&111011001000\\
&111010101000\\
&111010011000\\
&111001101000\\
&111001001000\\
&111100010000\\
&111100110000\\
&111101010000\\
&111101110000\\
&111110100000\\
\hline
&\vdots
\end{tabular}
}
\caption{First elements of $\mathcal W$.} \label{w}
\end{center}
\end{table}

\medskip

A more fruitful study can be based on the well-known combinatorial objects termed \emph{Dyck words} (see \cite{stanley} for further details). A Dyck word $v$
is a binary string composed of $n$ zeros and $n$ ones such that $|\alpha|_1 \geq |\alpha|_0$ for each prefix $\alpha$ of $v$, where $|\alpha|_1$ and $|\alpha|_0$ denote the number of ones and zeros in $\alpha$, respectively. By definition, a Dyck word necessarily starts with 1 and ends with 0. Denoting by $D_{2n}$ the set of all $2n$-length Dyck words, we have that the cardinality of
$D_{2n}$ is given by the $n$-th Catalan number $C_n=\frac{1}{n+1}{2n \choose n}$ whose generating function is $C(x)=\displaystyle\frac{1 - \sqrt{1 - 4x}}{2x}$ (see sequence A000108 in The On-line Encyclopedia of Integer Sequences).

Let $\mathcal D = \displaystyle \bigcup_{i\geq0}\{1\alpha0 : \alpha \in D_{2i}\}$ be the set constituted by the union of the binary words beginning with 1 linked to a $2i$-length Dyck word $\alpha$ and ending with 0, where $\alpha$ can be also empty. We have the following proposition.

\begin{proposition}
$\mathcal D$ is a cross-fix-free code.
\end{proposition}

\emph{Proof.} \quad Each $v \in \mathcal D$ can be factorized as $v=\alpha w \beta$ for some $\alpha, \beta \in \{0,1\}^+$ and $w \in \{0,1\}^*$. Since $|\alpha|_1 > |\alpha|_0$ and $|\beta|_1 < |\beta|_0$, then $\alpha \neq \beta$, for each $\alpha, \beta$ having the same length. So, $v$ is bifix-free. Analogously, given two distinct words $v,v' \in \mathcal D$, they can be written as $v=\alpha w$ and $v'=w'\beta'$, for some $\alpha,\beta',w, w' \in \{0,1\}^+$. Since, $|\alpha|_1 > |\alpha|_0$ and $|\beta'|_1 < |\beta'|_0$, then $\alpha \neq \beta'$, for each $\alpha, \beta$ having the same length. So, no (non-empty) proper prefix of $v$ is a suffix of $v'$, and vice versa, for each $v, v' \in \mathcal D$.

Now, we have to prove that $\mathcal D$ is also fix-free, that is, no codeword in the set is a prefix or a suffix of any other. Suppose ad absurdum that $v'= v \gamma$, with $\gamma \in \{0,1\}^+$, for some $v, v' \in \mathcal D$ such that $|v| < |v'|$. Being $v' \in \mathcal D$, then $|\alpha|_1 > |\alpha|_0$ for each prefix $\alpha$ of $v'$. Thus, we have a contradiction since $v \in \mathcal D$ and so $|v|_1=|v|_0$. Analogously, $v'\neq \gamma v$ for any $\gamma \in \{0,1\}^+$, hence $\mathcal D$ is a cross-fix-free code.
\cvd

For any $n \geq 2$, let $\mathcal D (n) = \displaystyle \bigcup_{i=0}^{\lfloor\frac{n-2}{2}\rfloor}\{1\alpha0 : \alpha \in D_{2i}\}$ be the subset of $\mathcal D$ containing binary strings having length at most $n$. Then,

$$|\mathcal D(n)|= \sum_ {i=0}^{\lfloor\frac{n-2}{2}\rfloor} C_i $$

\noindent and the generating function counting the size of $\mathcal D$ according to the semi-length of its codewords is

$$\mathcal D(x)= \displaystyle\frac{x \ C(x)}{1-x}.$$

The first numbers of $|\mathcal D(n)|$ are:
$$1,2,4,9,23,65,197,626,2056,6918,23714,82500,$$
for $n=2i$, with $i=1,2,3,\ldots, 12$, respectively. Obviously, if $n$ is odd then $\mathcal D(n)$ contains strings having length at most $n-1$.

In Table \ref{dyck} we list the first elements of the set $\mathcal D$ according to the length of the strings (in particular, the set $\mathcal D(10)$ is shown).

\begin{table}[h!]
\begin{center}
{\footnotesize
\begin{tabular}{lc}
n=2 & 10\\
\hline
n=4 & 1100\\
\hline
n=6 & 111000\\
&110100\\
\hline
n=8 & 11110000\\
&11101000\\
&11011000\\
&11100100\\
&11010100\\
\hline
n=10 & 1111100000\\
&1111010000\\
&1110110000\\
&1111001000\\
&1110101000\\
&1101110000\\
&1111000100\\
&1101101000\\
&1110100100\\
&1101011000\\
&1101100100\\
&1110010100\\
&1110011000\\
&1101010100\\
\hline
&\vdots
\end{tabular}
}
\caption{First elements of $\mathcal D$.} \label{dyck}
\end{center}
\end{table}

Note that, for any $n \geq 8$, $\mathcal W(n) \cap \mathcal D(n) \neq \emptyset$, and in general $\mathcal W(n)$ is not included in $\mathcal D(n)$.\\

Moreover, the code $\mathcal D'= \displaystyle \bigcup_{i \geq 0} \{11\alpha0 : \alpha \in D_{2i}\}$ can be also defined, obtaining a code having all its codewords of odd length.

\subsection{About the non-expandable codes}

As well as in the theory of cross-bifix-free codes, a crucial study concerns the construction
of cross-fix-free codes which are also non-expandable.

For this purpose, we extend the definition of non-expandable cross-bifix-free codes to cross-fix-free codes as follows.

\begin{definition}\label{expa}
Let $X$ be a cross-fix-free code and $X(n)$ be the subset of $X$ containing the strings having length at most $n$. For any fixed $n$, we say that $X(n)$ is \emph{non-expandable} if and only if for each binary bifix-free string $\psi$ having length at most $n$, with $\psi \notin X(n)$, we have that $X(n) \cup \psi$ is not a cross-fix-free code.
\end{definition}

For example, $\mathcal W(9)=\{11101000, 111001000, 111011000\}$ is an expandable cross-fix-free code, since it can be expanded by the set of binary strings $\{10,1100,110100\}$ maintaining the cross-fix-free property.
On the contrary, for the code $\mathcal D(n)$ we have the following proposition.

\begin{proposition}
For any fixed $n\geq 2$, $\mathcal D(n)$ is a non-expandable cross-fix-free code.
\end{proposition}

\emph{Proof.} \quad Fixed $n \geq 2$, let $\psi$ be a binary bifix-free string having length at most $n$ such that $\psi \notin \mathcal D(n)$, and
consider $T = \mathcal D(n) \cup \{\psi\}$.
If $\psi$ begins with $0$ (ends with 1) then $T$ is not cross-fix-free since any string in $\mathcal D(n)$ ends with $0$ (begins with 1). Consequently we have to consider $\psi$ as a string beginning with 1 and ending with $0$.

Let $h = |\psi|_1 - |\psi|_0$, we need to distinguish three different cases: $h = 0$, $h > 0$ and $h < 0$.
\begin{itemize}
\item[$\bullet$]
If $h =0$, then $\psi$ admits a (non-empty) proper prefix $u$ having length at most $n-2$ such that $|u|_0=|u|_1$, otherwise if such a prefix $u$ would not exist, then $\psi$ should be in $\mathcal D(n)$ against our assumption.  Let $u$ be the smallest proper prefix $u$ of $\psi$ such that $|u|_0=|u|_1$, then $u=1 \alpha 0$ where $\alpha$ is a Dyck word which can be also empty. So, $u \in \mathcal D(n)$ and $T$ is not a cross-fix-free code.

\item[$\bullet$]
If $h > 0$, $\psi$ can be written as
\[ \psi = \phi \; 1 \; \mu_1 \; 1 \; \mu_2 \; \cdots \; 1 \; \mu_h, \]
where $\phi$ is a word (possibly empty) satisfying $|\phi|_1 = |\phi|_0$ and beginning with $1$, and $\mu_1, \dots, \mu_h$ are Dyck words with $\mu_h$ non-empty as $\psi$ ends with $0$. Since $\mu_h$ is a Dyck word having length at most $n-1$, then it can be factorized as $\mu_h=\beta'1\alpha'0$, where $\beta',\alpha'$ are two Dyck words (possibly empty). So, the suffix $1\alpha'0$ of $\psi$ is a word in $\mathcal D(n)$ and $T$ is not a cross-fix-free code.

\item[$\bullet$]
If $h < 0$, $\psi$ can be written as
\[ \psi = \mu_{h} \; 0 \; \cdots \; \mu_2 \; 0 \; \mu_1 \; 0 \; \phi \]
where $\phi$ is a word (possibly empty) satisfying $|\phi|_1 = |\phi|_0$ and ending with $0$, and $\mu_1, \dots, \mu_{h}$ are Dyck words with $\mu_{h}$ non-empty as $\psi$ begins with 1. Since $\mu_{h}$ is a Dyck word having length at most $n-1$, then it can be factorized as $\mu_h= 1\alpha'0 \beta'$, where $\alpha',\beta'$ are two Dyck words (possibly empty). So, the prefix $1\alpha'0$ of $\psi$ is a word in $\mathcal D(n)$ and $T$ is not a cross-fix-free code.

\end{itemize}

We proved that $\mathcal D(n)$ is a non-expandable cross-fix-free code, for any $n \ge 2$.
\cvd

Definition \ref{expa} can be easily adapted to the strong cross-fix-free codes. Unfortunately, for any fixed $k \geq 3$, the code $\mathcal V^{(k)}(n)$ presented in Section 2 is an expandable strong cross-fix-free code, since for instance the string $11(01)^{\lfloor\frac{n-4}{2}\rfloor}00$ can be added to $\mathcal V^{(k)}(n)$ maintaining the strong cross-fix-free property (this is also valid if we consider $\mathcal V^{(k)}(n) \cup \{1^k0^k\}$). A more deep inspection is needed in this direction, and a further line of research could take into consideration strong cross-fix-free codes which are non-expandable.


\end{document}